\documentclass{article}
\newcommand{\cosec}{{\rm cosec}}

\begin{document}
\Large
{\begin{center}  \textit{Ab Initio} Method for Obtaining Exactly Solvable Quantum Mechanical Potentials\end{center} %
\large \begin{center} Asim Gangopadhyaya\footnote{agangop@luc.edu, asim@uic.edu}, Jeffry V. Mallow%
 \footnote{jmallow@luc.edu}\\
Department of Physics, Loyola University Chicago,  \\Chicago,
Illinois 60626  \vspace*{.25in}\end{center}%
}
\abstract{The shape invariance condition is the integrability condition in supersymmetric quantum mechanics (SUSYQM). It is a difference-differential equation connecting the superpotential W and its derivative at two different values of parameters.  We show that this difference equation is equivalent to a non-linear partial differential equation whose solutions are translational shape invariant superpotentials. In lieu of trial and error, this  method provides the first \textit{ab initio} technique for generating shape invariant superpotentials.}

\vspace*{.2in}

\noindent \normalsize Supersymmetric quantum mechanics (SUSYQM)\cite{SUSYQM} extends Dirac's factorization method%
\footnote{Dirac however, credits Fock with that discovery: P. A. M. Dirac, Communications of the Dublin Institute of Advanced Study, A1, pp. 5-7 (1943); Fock, Zeits. f. Phys. 49 339 (1928).} %
for the harmonic oscillator to a large number of other potentials, whose energy eigenvalues can then  be obtained algebraically. The SUSYQM extension consists of introducing a superpotential $W(x,a)$ that generates two partner Hamiltonians, both with the same energy eigenvalues\footnote{One of the partner potentials has an `extra' zero-energy ground state.}. These  partner Hamiltonians are given by %
\begin{eqnarray} H_{\mp} &=& A^{\pm}\, A^{\mp} = \left(\mp\, \hbar \, {d \over {dx}} +W(x,a)\right) \left(\pm \,\hbar \, {d \over {dx}} + W(x,a) \right)\nonumber
\\
&& \nonumber
\\
&=& -\,\hbar^2\,{{d^2} \over {dx^2}}+ W^2(x,a)\mp \hbar \,
{{dW(x,a)} \over
{dx}}
 = -\,\hbar^2\,{{d^2} \over {dx^2}}+ V_{\mp}(x,a),
\label{Schrodinger}
\end{eqnarray}
where partner potentials $V_{\pm}(x,a)$ are related to superpotential W by $W^2(x,a)\pm \hbar \,
{{dW(x,a)} \over
{dx}}$. (We have set $2m=1$.) One of the best known examples is the superpotential $W(x,a) = -a \cot x$ with $a>0$, and $0 <x < \pi$. The partner potentials generated by this $W(x,a)$ are: %
 $V_\pm (x,a) = W^2(x,a) \pm  \hbar \, \frac{dW(x,a)}{dx} =
a(a\pm\hbar \, )
~\cosec^2x - a^2.$
For the special case of $a=\hbar $, the
potential $V_-(x,\hbar )$ is a constant function: $-\, a^2=-\,
\hbar^2$; i.e., just an infinite one-dimensional square well
potential whose bottom is set to $- \hbar^2$, while its partner
potential $V_+(x,\hbar)$ is given by $\hbar^2\,(2\cosec^2x - 1)$.
Thus, the  much less familiar $\cosec^2$ potential has the same
spectrum as the well-known infinite well
(except for the infinite well's `extra' ground state).

\vspace*{.15in}
However, for any pair of partner potentials, one would still have to know the spectrum of one of the partners to obtain the other.  The critical advance in SUSYQM was the discovery
\cite{ShapeInvariance1} and subsequent rediscovery of shape invariance (SI) \cite{ShapeInvariance2}. A system is called shape invariant if partner potentials could be shown to differ only by the value of a
parameter and an additive constant:
 \begin{eqnarray}
V_+(x,a_0)~=~V_-(x,a_1)+  R(a_0)  ~, %
\label{SIeq}\end{eqnarray}
in which case each spectrum can be generated without reference to its partner.

We can write, more symmetrically %
\begin{eqnarray}
V_+(x,a_0)+
g(a_0)~=~V_-(x,a_1)+  g(a_1)~, %
\nonumber\end{eqnarray}
where $R(a_0) \equiv g(a_1)-g(a_0)$.  That is, \begin{eqnarray}
W^2(x,a_0) + \hbar \, \frac{d W(x,a_0)}{d x} + g(a_0)~=~W^2(x,a_1)
- \hbar \,
\frac{d W(x,a_1)}{d x} + g(a_1) %
\label{SI}\end{eqnarray}
The shape invariance condition in the form of this difference-differential equation has also been studied for dynamical systems, where it is known as the infinite-dimensional dressing \cite{Shabat}.
In the case of our example above, $W(x,a) =
-\,a \cot x$, the partner potentials are related
by
\begin{eqnarray}
V_+(x,a)+a^2&=& V_-(x,a+\hbar) +  (a+\hbar)^2 . \label{eq3}
\end{eqnarray}
Thus they are shape invariant: a shift of parameter $a_0\equiv a$
to $a_1\equiv a+\hbar$ $g(a_0)=a_0^2$, and an additional constant
$(a+\hbar)^2-a^2$ is all that distinguishes the potentials
$V_+(x,a)$ and $V_-(x,a+\hbar)$ from each other.  But we could just as
well have chosen $a_0\equiv a+\hbar$ and $a_1\equiv a+2 \hbar$, and written %
\begin{eqnarray}%
V_+(x,a+\hbar) + (a+\hbar)^2 &=& V_-(x,a+2\hbar) + (a+2\hbar)^2~.%
\label{eq4} \end{eqnarray}%
By induction, we can generate a family of partner potentials
related by $a_n = a_{n-1}+\hbar$. This form of shape invariance is called ``additive" or
``translational" \cite{SUSYQM}.  This will be the case
for all the shape invariant superpotentials we shall consider in this paper. (There are other forms of shape invariance, but we shall not consider them here.)  We emphasize that the shape-invariance condition (Eq. (\ref{SI})) is non-linear in $W$; therefore one cannot simply add arbitrary constants to $W$ and maintain shape-invariance. This extremely strong constraint restricts the known shape invariant superpotentials to a very small number.

For shape invariant superpotentials,  SUSYQM yields
\cite{SUSYQM} energy eigenvalues $$ ~E^{(-)}_n(a_0)=
\Sigma_{i=0}^{n-1} R(a_i) = g(a_n) - g(a_0).$$  Thus, for the
infinite well spectrum, with its bottom at $-a^2=-\hbar^2$,
\\ \vspace*{-.25in}\begin{center} $E^{(-)}_n(a_0) = g(a_n) - g(a_0) = ( n+1)^2\, \hbar^2 -
\hbar^2\,$ \footnote{N.B. Unlike the standard convention in
quantum mechanics texts, $n$ begins at $0$, the well width is
$\pi$, and the well bottom is at $-\,\hbar^2$. This results into a
ground state energy $E^{(-)}_0 = 0$, a requirement of the SUSYQM
formalism.}.\end{center}

\noindent
Note that Eq. (\ref{SI}) a
difference-differential equation; that is,  it relates the square
of the superpotential $W$ and its spatial derivative computed at
two different  parameter values: $(x,a_0\equiv a)$ and $(x,a_1
\equiv a+ \hbar)$. Our objective
is to show that this integrability condition can be
expressed as a non-linear partial differential equation that is
local in nature; i.e., all terms in the equation can be computed
at the same point $(x,a)$. This has the obvious advantage of
mathematical familiarity (at least to physicists).  It
provides a  systematic method for finding the superpotential, and thus
$g(a_n)$ and from it the energy spectrum of translational shape
invariant superpotentials.  We shall show that all known translational
shape invariant superpotentials \cite{SUSYQM} are its solutions.

\vspace*{.15in}\noindent
As in the above example we will only consider translational shape invariance, where the parameters are related by $a_1 = a_0 +\hbar$. For some superpotentials, such as Morse, successive parameters see a decrease by $\hbar$ instead of an increase. Such superpotentials are also included in these discussions. For the Morse superpotential $W(x,A)  = A-Be^{-x}$, if we insist that $A\rightarrow A-\hbar$ at each step of change of parameters, $A$ of Morse will be related to the $a$ parameter by $A= - a$.
Note that we have made no assumptions about $\hbar$: for now, it is simply a
part of the parameter defining $W$. Since this equation is valid
for any value of $\hbar$, we will treat $x$ and $a -
\hbar$ on equal footing as independent variables.
Equation (\ref{SI}) can now be written as:%
\begin{eqnarray}
W^2(x,a- \hbar) + \hbar \, \frac{\partial W(x,%
a- \hbar)}{\partial x} + g(a- \hbar)~=~W^2(x,a) -
\hbar \, \frac{\partial W(x,a)}{\partial x} + g(a)\label{SIPDE}\end{eqnarray}
where $a_i = a$ and $a_{i-1} = a- \,\hbar$. If we now differentiate Eq. (\ref{SIPDE})
with respect to $\hbar$, noting that $W^2(x,a)$ and $g(a)$ are independent of $\hbar$, we obtain
\begin{eqnarray}
2 \,W(x,a- \hbar \,)\,\frac{\partial W(x,a- \hbar)}{\partial \hbar} + \frac{\partial W(x,a-\hbar)}{\partial x} + \hbar \, \frac{\partial}{\partial \hbar }\frac{\partial W(x,a- \hbar)}{\partial x}+ \frac{\partial g(a- \hbar)}{\partial \hbar}~= - \frac{\partial W(x,a)}{\partial x}
\label{derivSIPDEwrthbar}
\end{eqnarray}

\vspace*{.15in}\noindent
Now because of translational shape
invariance, the dependence of $W$ on $a$ and $\hbar$ is through the
linear combination $a- \hbar$; therefore, the derivatives
of $W$ with respect to $a$ and $\hbar$ are related by: $ \frac{\partial W(x,a- \hbar)}{\partial \hbar} = -\,\frac{\partial W(x,a- \hbar)}{\partial a}$.

Then we may rewrite Eq. (\ref{derivSIPDEwrthbar}):
\begin{eqnarray}
- 2  W(x,a- \hbar)\frac{\partial W(x,a-
\hbar)}{\partial a} + \frac{\partial W(x,a-
\hbar)}{\partial x} -  \hbar \, \frac{\partial}{\partial a
}\frac{\partial W(x,a- \hbar)}{\partial x}-
\frac{\partial g(a- \hbar)}{\partial a}~=\nonumber~ -
\frac{\partial W(x,a)}{\partial x}
\end{eqnarray}
But this must be true for any value of $\hbar$.
If we then set $\hbar=0$, shape invariance implies the following
condition for the superpotential $W(x,a)$:
\begin{eqnarray}%
2 \,  \, W(x,a) \, \left( \frac{\partial W(x,a)}{\partial
a}\right) - 2\, \frac{\partial W(x,a)}{\partial x} +  \,
\frac{\partial g(a)}{\partial a} = 0
\label{PDE} \end{eqnarray}
This is the desired partial differential equation which is equivalent to the original difference-differential equation for shape invariance.
Thus, we have shown that all shape invariant superpotentials with $a_{i+1} = a_i+\hbar$ are solutions of the above non-linear partial differential equation. Before we develop a method to determine solutions of this complex equation; i.e., to find shape invariant superpotentials, we will as an example show that two known shape invariant superpotentials are indeed solutions of this differential equation, and we shall obtain their spectra. It can be easily verified that all members of the list given in Ref. \cite{SUSYQM,Dutt} are solution of this equation.

\noindent{\bf Morse potential:} The superpotential  is
given
by %
\begin{equation}W(x,A) = A-Be^{-x} \end{equation} %
Substituting this superpotential in Eq. (\ref{PDE}), we get:%
\begin{eqnarray}%
2A +\frac{\partial g(A)}{\partial A} =0~,%
\end{eqnarray}
whence, $A = - \frac 1 2 \,
\frac{\partial g(A)}{\partial A}$. As stated earlier, we have $a_0 = - A$
and $a_n = a_0+ n\,\hbar = -A + n\,\hbar$, with $g(A)= -A^2$. This leads to $E_n =  A^2 - (A-n\,\hbar)^2$.

\vspace*{.25in}\noindent
{\bf {Scarf I potential}:} The superpotential is
given by %
\begin{equation}W(x,A) = A \tan\,x + B \sec\,x ~~~~~A>B, \end{equation} where $A$ is the parameter
that changes. Substituting this superpotential into Eq.
(\ref{PDE}), we get:
\begin{eqnarray}%
2 \,   \left( A \tan\,x + B \sec\,x  \right) %
\left( \tan\,x  \right)
&-& 2\, \left( A \sec^2\,x + B \sec\,x\tan\,x  \right) +  %
 \, \frac{\partial g(A)}{\partial A}=0\nonumber\\
\nonumber\\
 \, \frac{\partial g(A)}{\partial A} - 2\,A
&=&0\nonumber ~~,%
\end{eqnarray}
hence, $g(A)= A^2$.   Thus in our
notation $a_0=A$ and $a_n = a_0 + n\, \hbar = A+n\,\hbar$, and
$E_n = g(a_n) - g(a_0) = (A+n\,\hbar)^2 - A^2$.

\vspace*{.15in}\noindent
One can check that all examples listed in Ref. \cite{SUSYQM,Dutt}  satisfy the partial differential equation given in Eq.
(\ref{PDE}).  For a given shape-invariant superpotential,  Eq. (\ref{PDE}) determines $g(a)$, and hence
its eigenspectrum can be obtained using $E_n = g(a_n) - g(a_0)$.

However, our true goal is the opposite.  Until now, finding shape invariant superpotentials was left to
excellent intuitions or trial and error. We develop for the first time an \textit{ab initio} systematic method for these superpotentials.  The remainder of this paper shows how this can be done.

Since all translation shape invariant superpotentials are necessarily solutions of Eq. (\ref{PDE}), finding solutions of this equation will help us in our quest for shape invariant superpotentials. For this, we
have developed a method, similar to  the separation of variables, to solve this non-linear partial differential equation. Using our method, we are able to derive all the known translational shape invariant superpotentials listed in Ref. \cite{SUSYQM,Dutt}. The method also provides a check as to whether there are additional, as-yet-undiscovered superpotentials.

At this point, it is tempting to solve Eq. (\ref{PDE})  by assuming various forms for the term $\frac{\partial g(a)}{\partial a}$. However, instead of that rather random process, we choose to try a factorizable ansatz
\begin{eqnarray}%
W(x,a) = \sum_{i=1}^N {\cal{A}}_i (a)\, {\cal{X}}_i(x)
\label{Ansatz} \end{eqnarray}
\normalsize where ${\cal{A}}_i (a)$ and ${\cal{X}}_i (x)$ are assumed
to be respectively functions of $a$ and $x$. Thus, to determine
$W(x,a)$, we will need to find $2N$ functions ${\cal{A}}_i (a)$
and ${\cal{X}}_i (x)$. Eq. (\ref{PDE}) provides one equation
involving these functions. So, we need another $ 2N-1$
constraints to fully determine $W(x,a)$. We will consider the
above ansatz for various values of $N$.  Throughout this work, we shall use the following notation:
\begin{itemize}
\item
Lower case Greek letters will denote $a$- and $x$- independent constants;
\item
Upper case Latin letters with $a$ in parentheses  will denote $a$-dependent, but $x$-independent ``constants."
\end{itemize}

\noindent
{\bf \textit{N}=1 - Single-Term Superpotentials:}
\begin{eqnarray}%
W(x,a) = {\cal{A}}(a)\, {\cal{X}}(x)
\label{Ansatz1} \nonumber\end{eqnarray}
\normalsize Substituting this ansatz into Eq.(\ref{PDE}) leads to

\begin{eqnarray}%
2 \,  \, \left(  {\cal{A}}(a)\, {\cal{X}}(x)  \right) \,
\left( {\cal{X}}(x)\,\frac{d\,{\cal{A}}(a)}{d a}\right) - 2\,
 {\cal{A}}(a)\, \frac{d\, {\cal{X}}(x) }{d x} +  \,
 \frac{d
g(a)}{d a} = 0 \nonumber\\
\nonumber\\
{\rm Suppressing~the}~x~{\rm and}~a~{\rm dependence:}~~2 \, {\cal{A}}\,   \, %
\underbrace{%
 \left(  {\cal{X}}^2\,\frac{d\,{\cal{A}}}{d a}
-   \frac{d\, {\cal{X}}}{d x}\right) %
}_{H(a)}%
+ \, \,
 \frac{d
g}{d a} = 0 %
\label{NPDE1} %
\end{eqnarray}%
\normalsize In Eq.(\ref{NPDE1}), we see that the expression $
\left( \,  {\cal{X}}^2\,\frac{d\,{\cal{A}}}{d a}
-   \frac{d\, {\cal{X}}}{d x}\right) %
$ can at most be a function of the parameter $a$, and hence, we
set it equal to ${H(a)}$.  For $H(a) \neq 0$~\footnote{For $H(a) = 0$, the solution of the above differential equation is ${\cal{X}} \sim  1/x$ which is in fact a case of broken SUSY, and holds no bound states.}, this leads to
\begin{equation}{\cal{X}}= -
~ \sqrt%
{%
\frac{H}{ \frac {d\,{\cal{A}}}{da} } %
}%
\tanh \left(
{{\sqrt {H\,{\frac {d{\cal{A}}}{da}} } } }
~\left( x-x_{{0}} \right)  \right)
\label{Solution1}\end{equation} %
\vspace*{.15in}

\normalsize
\noindent
Since, ${\cal{X}}(x)$ is independent of $a$, we must have, %
${H \left( a \right)}\sim { \frac {d\,{\cal{A}}(a)}{da} }$ and %
${H \left( a \right)}\sim \left( \frac
{d\,{\cal{A}}(a)}{da}\right)^{-1} $. These two conditions can only be met if ${H \left( a
\right)}$ and ${ \frac {d\,{\cal{A}}(a)}{da} }$ are constants,
independent of the parameter $a$.  Thus ${\cal{A}}$ must be a
linear function of $a$: ${\cal{A}}= \mu a+ \beta,$ and
${H \left( a \right)}=\gamma$ where the Greek-letter constants are
all $a$-independent.

This leads to the superpotential  $W(x,a) = -~ (\mu a+ \beta)\sqrt%
{%
\frac{ \gamma}{ \mu } %
}
\tanh \left(
{{\sqrt { \gamma  \,\mu } } }
~\left( x-x_{{0}} \right)  \right)
$.
By a scaling ($\gamma \mu = 1$) and a shift ($x_0=0$) of $x$, we can write giving
\begin{equation}
  W(x,a) = A\, \tanh  x~,
\end{equation}
where $A=\left(\frac{\beta}{\mu}-a\right) $. This superpotential represents unbroken SUSY and holds finite number of bound state. It is a special case of the Scarf II and the Rosen-Morse superpotentials  \cite{SUSYQM,Dutt} .
Note that, for negative values of $H$, instead of the hyperbolic trigonometric function, we get the superpotential $W(x,a) = a \tan x~.$  Note also that this formulation includes terms like $\cot$ and $\coth$ as solutions of the same differential equation for different values of $H(a)$ and $ \frac {d{\cal{A}}}{da} $.  For ${\cal{A}} = $ constant, Eq. (\ref{NPDE1}) generates  %
\begin{equation}W(x,a) = \frac {\omega} 2 \,x~~,\end{equation}
the superpotential for the \emph{harmonic oscillator}, where we have set $H(a)=\,-\,\frac 1
2 \omega$.

This exhausts the solutions for the $N=1$ ansatz.  Note that by definition it can only give one of the terms in any superpotential.  
  \vspace*{.25in}

\newpage
\noindent
{\bf \textit{N}=2 - Two-Term Superpotentials:}
\begin{eqnarray}%
    W(x,a) &=& \sum_{i=1}^2 {\cal{A}}_i (a)\, {\cal{X}}_i(x) \nonumber \\
    &=& {\cal{A}}_1 (a)\, {\cal{X}}_1(x) + {\cal{A}}_2 (a)\,
    {\cal{X}}_2(x)
    \label{Ansatz2} \nonumber%
\end{eqnarray}%
\emph{We would like to very strongly emphasize that neither the functions ${\cal{A}}_1(a)$ and  ${\cal{A}}_2 (a)$, nor  ${\cal{X}}_1(x)$ and  ${\cal{X}}_2(x)$, should be linearly proportional to
each other, otherwise this case will reduce to
the case of $N=1$ considered above.}

\vspace*{.25in} \noindent Substituting the above ansatz in Eq.(\ref{PDE}) leads to
\begin{eqnarray}%
    2 %
    \left(  {\cal{A}}_1 \, {\cal{X}}_1 + {\cal{A}}_2
    \, {\cal{X}}_2   \right)
    \left(  {\cal{X}}_1 \, \frac{ d \, {\cal{A}}_1}{da} \, +
    {\cal{X}}_2 \, \frac{ d \, {\cal{A}}_2}{da}  \right)
    +  \, \frac{d g(a)}{d a} %
    =\left(  {\cal{A}}_1 \, \frac{ d \, {\cal{X}}_1}{dx} \, +
    {\cal{A}}_2 \, \frac{ d \, {\cal{X}}_2}{dx}  \right) \nonumber%
\end{eqnarray}%
\normalsize Expanding it, we get:
\begin{eqnarray}
   \underbrace{2{\cal{A}}_1 \, \frac{ d \, {\cal{A}}_1}{da}\,   {\cal{X}}_1 ^{\,2} }_{\rm Term \# 1}
\underbrace{ - 2{\cal{A}}_1  \frac{ d \, {\cal{X}}_1}{dx}}_{\rm Term \# 2}
 \underbrace{+ 2 {\cal{A}}_2 \, \frac{ d \, {\cal{A}}_2}{da} \,   {\cal{X}}_2 ^{\,2}}_{\rm Term \# 3}
 \underbrace{-  2 {\cal{A}}_2  \frac{ d \, {\cal{X}}_2}{dx}}_{\rm Term \# 4}
\underbrace{+   2\, {\cal{X}}_1 {\cal{X}}_2 %
      \left( {\cal{A}}_1 \frac{ d \, {\cal{A}}_2}{da} +
             {\cal{A}}_2 \frac{ d \, {\cal{A}}_1}{da}
      \right)}_{\rm Term \# 5}
+  \underbrace{ \, \frac{d g(a)}{d a}}_{\rm Term \# 6}
  = 0~.%
\label{coupled} \end{eqnarray}
This is of the form
$$\sum_i^6 F_i(a) G_i(x)=0$$
where
\begin{equation}
\begin{array}{ll}
F_1 \equiv 2\,   {\cal{A}}_1 \, \frac{ d \, {\cal{A}}_1}{da};~~~~~~~~~~~~~~~~~~~~~~~~~~~~
&G_1 \equiv \left( {\cal{X}}_1 \right)^{2}\\
F_2 \equiv -\,  2 {\cal{A}}_1   ;
&G_2 \equiv \frac{ d \, {\cal{X}}_1}{dx} \\%
F_3 \equiv 2\,   {\cal{A}}_2 \, \frac{ d \, {\cal{A}}_2}{da} ;
&G_3 \equiv \left( {\cal{X}}_2 \right)^{2}\\%
F_4 \equiv -\,  2 {\cal{A}}_2   ;
&G_4 \equiv \frac{ d \, {\cal{X}}_2}{dx} \\%
F_5 \equiv 2\,    \left( {\cal{A}}_1 \frac{ d \, {\cal{A}}_2}{da} +
             {\cal{A}}_2 \frac{ d \, {\cal{A}}_1}{da}
      \right) ;
&G_5 \equiv {\cal{X}}_1 {\cal{X}}_2\\%
F_6 \equiv  dg/da   ;
&G_6 \equiv 1
\end{array}
\end{equation}
\vspace*{0.2in}

This linear combination of various functions of $x$ must yield us functions $ {\cal{X}}_1$ and $ {\cal{X}}_2$ that are independent of $a$. This severely constrains the ${\cal{A}}_i$'s and $ {\cal{X}}_i$'s.  In fact, in the appendix we prove the following theorem:
\emph{If a combination of $F_i\,G_i$ is irreducibly constant, i.e., no smaller combination is independent of $x$, then all $F_i$'s must be proportional to each other.}
We will be making extensive use of this property throughout this paper, where we utilize the above constraints to obtain all two-term superpotentials.

In the notation we will be using, we will refer to $F_i\,G_i$ as the $i$-th term; e.g, ${2{\cal{A}}_1 \, \frac{ d \, {\cal{A}}_1}{da}\,   {\cal{X}}_1 ^{\,2} }$ as  {term \# 1}, ${2{\cal{A}}_1  \frac{ d \, {\cal{X}}_1}{dx}}$ as {term \# 2}, etc., as explicitly shown in Eq. (\ref{coupled}).
Of these terms, the sum of the first five terms is $x$-independent since the sixth term, $ \, \frac{d g(a)}{d a}$ clearly does not depend on $x$. However, a subcollection of the first five terms may also add up to a constant. If, for example, terms 2, 3 and 4 add up to a constant and no smaller group of them is a constant, we would call such a group irreducibly constant and denote it by \{2,3,4\}.

\newpage
\noindent
Our general method will be driven by the following constraints:
\begin{itemize}
\item
Any irreducible term or set of terms that is irreducible must be at most only $a$-dependent.
\item
In any irreducibly constant subgroup, the functions of ${\cal{A}}_i$ must be proportional to each other, as proved in the appendix.
\item
For those cases where only a single term is irreducible, then either the function of ${\cal{X}}_i $ must be a constant, or if it is not, then the function of ${\cal{A}}_i$ must be zero, to preserve the dependence on $a$ alone.
\end{itemize}

\begin{enumerate}
\item
Let us first consider the case that the sum $\sum_i^5 F_i(a) G_i(x)$ is irreducible; that is, no smaller group of terms is constant. We will denote this group by  \{1,2,3,4,5\}. In this case all $F_i$ are proportional to each other, as proved in the appendix. The proportionality of $F_2$ and $F_4$
implies that the ${\cal{A}}_i$'s are proportional and hence this reduces to the $N=1$ case.  We would like to emphasize this feature. \emph{Whenever an irreducible group contains both term \#2 and term \#4, we have $ {\cal{A}}_1 \sim  {\cal{A}}_2$, and hence a case of $N=1$.}

\item
Now let us consider the case where there are four-term combinations independent of $x$. This also implies that there is one $x$-independent term, which leads to severe constraint on $
{\cal{X}}_i$'s.  The remaining 4-term group has all corresponding
$F_i$'s proportional to each other.

There are five such possibilities, of which, due to symmetry between the first and third terms and between the second and fourth, two of the terms are similar to others. For example, the
condition ``${\cal{X}}_1=$constant" is equivalent to
``${\cal{X}}_2=$constant."~\footnote{As we did in the $N=1$ case, we must distinguish between constants that depend on $a$, and those that are  $a$- and $x$-independent constants.}

This leads to three cases:
\begin{enumerate}
\item  { \{2,3,4,5\}} + \{1\}
\begin{eqnarray}
 -\,  2{\cal{A}}_1  \frac{ d \, {\cal{X}}_1}{dx} \,
+    2 {\cal{A}}_2 \, \frac{ d \, {\cal{A}}_2}{da} \,   {\cal{X}}_2 ^{\,2} \,
 -\, 2 {\cal{A}}_2  \frac{ d \, {\cal{X}}_2}{dx} \,
  + 2\, {\cal{X}}_1 {\cal{X}}_2 %
      \left( {\cal{A}}_1 \frac{ d \, {\cal{A}}_2}{da} +
             {\cal{A}}_2 \frac{ d \, {\cal{A}}_
             1}{da}
      \right){\rm is ~ an ~}a{\rm-dependent ~constant}.
\label{4Term2345}\nonumber
\end{eqnarray}
Since terms \#2 and \#4 appear in the same irreducible combination, the problem reduces to the $N=1$ case.

\item  { \{1,3,4,5\}} + \{2\}
\begin{eqnarray}
 2{\cal{A}}_1 \, \frac{ d \, {\cal{A}}_1}{da}\,   {\cal{X}}_1 ^{\,2}  \,
+    2 {\cal{A}}_2 \, \frac{ d \, {\cal{A}}_2}{da} \,   {\cal{X}}_2 ^{\,2} \,
 -\, 2 {\cal{A}}_2  \frac{ d \, {\cal{X}}_2}{dx} \,
  + 2\, {\cal{X}}_1 {\cal{X}}_2 %
      \left( {\cal{A}}_1 \frac{ d \, {\cal{A}}_2}{da} +
             {\cal{A}}_2 \frac{ d \, {\cal{A}}_
             1}{da}
      \right)   \equiv H_1(a)\label{4Term1345} \end{eqnarray}
and $\left[ 2{\cal{A}}_1  \frac{ d \, {\cal{X}}_1}{dx}\right] = -  \, \frac{d g(a)}{d a} - H_1(a)$. %
From the irredicibility of { \{1,3,4,5\}}, we have:
\begin{eqnarray}
    {\cal{A}}_1 \, \frac{ d \, {\cal{A}}_1}{da} \,
\sim
    {\cal{A}}_2 \,  \frac{ d \, {\cal{A}}_2}{da} \,
\sim
        {\cal{A}}_2
\sim
      \left( {\cal{A}}_1 \frac{ d \, {\cal{A}}_2}{da} +
             {\cal{A}}_2 \frac{ d \, {\cal{A}}_1}{da}
      \right)
\sim
    H(a)~.\nonumber
\end{eqnarray}
From
${\cal{A}}_2 \,  \frac{ d \, {\cal{A}}_2}{da} \, \sim  {\cal{A}}_2$, we get ${\cal{A}}_2= \alpha a+\beta.$
From
$    {\cal{A}}_1 \, \frac{ d \, {\cal{A}}_1}{da} \,
\sim
    {\cal{A}}_2 \,  \frac{ d \, {\cal{A}}_2}{da} \,,
$
we get $\left( {\cal{A}}_1\right)^2= \left( {\cal{A}}_2\right)^2+\delta = \left( \alpha a+\beta\right)^2+\delta.$
However, from
$     \left( {\cal{A}}_1 \frac{ d \, {\cal{A}}_2}{da} +
            {\cal{A}}_2 \frac{ d \, {\cal{A}}_1}{da}
       \right)
    \sim  {\cal{A}}_2 = (\alpha a + \beta)$, we get ${\cal{A}}_1 = \gamma +\zeta \exp\left(-a^2/2 - (\beta\,a)/\alpha    \right)$. These two expressions for
${\cal{A}}_1$ are consistent only if it is a constant independent
of $a$.  But $\left( {\cal{A}}_2\right)^2= \left( {\cal{A}}_1\right)^2-\delta$, so ${\cal{A}}_2$ is constant as well. Constancy of both ${\cal{A}}_1$ and ${\cal{A}}_2$ is a special case of ${\cal{A}}_1\sim {\cal{A}}_2$. Thus, this problem reduces to the $N=1$ case.
\item  { \{1,2,3,4\}} + \{5\}
\begin{eqnarray}
 2{\cal{A}}_1 \, \frac{ d \, {\cal{A}}_1}{da}\,   {\cal{X}}_1 ^{\,2}  \,
-\,  2{\cal{A}}_1  \frac{ d \, {\cal{X}}_1}{dx} \,
+    2 {\cal{A}}_2 \, \frac{ d \, {\cal{A}}_2}{da} \,   {\cal{X}}_2 ^{\,2} \,
 -\, 2 {\cal{A}}_2  \frac{ d \, {\cal{X}}_2}{dx} \,
 \equiv H(a)~,
\label{4Term1234} \end{eqnarray}
and $\left[ 2\, {\cal{X}}_1 {\cal{X}}_2 %
                 \left( {\cal{A}}_1 \frac{ d \, {\cal{A}}_2}{da} +
                 {\cal{A}}_2 \frac{ d \, {\cal{A}}_1}{da}
                 \right)\right]=  -  \, \frac{d g(a)}{d a} - H(a) $.
From the second and fourth terms of Eq. (\ref{4Term1234}), the irreducibility of { \{1,2,3,4\}}  implies that $ {\cal{A}}_1 \sim  {\cal{A}}_2$. Thus problem reduces to the $N=1$ case as well.
 \end{enumerate}

\item
There are ten possible ways to set three-term groups equal to a
constant. However, due to the symmetry of terms with respect to ${\cal{X}}_1$ and ${\cal{X}}_2$, we see that there are only six combinations we need
to consider. We list them all below. They are:
\begin{enumerate}
\item
Terms $\{1,2,3\}$ add up to an $a$-dependent constant.\\ Let us first consider
 { \{1,2,3\}}+\{4,5\}.
\begin{eqnarray}  
 2{\cal{A}}_1 \, \frac{ d \, {\cal{A}}_1}{da}\,   {\cal{X}}_1 ^{\,2}  \,
 -\,  2{\cal{A}}_1  \frac{ d \, {\cal{X}}_1}{dx} \,
+    2 {\cal{A}}_2 \, \frac{ d \, {\cal{A}}_2}{da} \,   {\cal{X}}_2 ^{\,2} \,
= H(a).  %
\end{eqnarray}
Since we have %
$    {\cal{A}}_1 \frac{ d \, {\cal{A}}_1}{da} \, \sim
{\cal{A}}_1 \sim \,  \, {\cal{A}}_2 \frac{ d \,
{\cal{A}}_2}{da} , $ we find from the first condition that ${\cal{A}}_1$ must be a linear
function of $a$, i.e.,   ${\cal{A}}_1= \alpha \,a+\beta$, and from the second condition
that $\left(  {\cal{A}}_2 \right)^2 = \gamma \left(  {\cal{A}}_1 \right)^2 +\delta =\gamma (\alpha \,a+\beta)^2 +\delta$. From the constancy of \{4,5\}, we have
\begin{eqnarray}
 -\, 2 {\cal{A}}_2  \frac{ d \, {\cal{X}}_2}{dx} \,
  + 2\, {\cal{X}}_1 {\cal{X}}_2 %
      \left( {\cal{A}}_1 \frac{ d \, {\cal{A}}_2}{da} +
             {\cal{A}}_2 \frac{ d \, {\cal{A}}_
             1}{da}
      \right)
= J(a)~.
\end{eqnarray}
This gives
${\cal{A}}_2 \sim       \left( {\cal{A}}_1 \frac{ d \, {\cal{A}}_2}{da} +
             {\cal{A}}_2 \frac{ d \, {\cal{A}}_
             1}{da}
      \right)   \equiv  \kappa\,
      \left( {\cal{A}}_1 \frac{ d \, {\cal{A}}_2}{da} +
             {\cal{A}}_2 \frac{ d \, {\cal{A}}_
             1}{da}
      \right)   .$  %
       Substituting ${\cal{A}}_1 = (\alpha a + \beta)$, we get the following differential equation:
\mbox{$\kappa  \left( \alpha a + \beta \right) \frac{ d \, {\cal{A}}_2}{da} +  \left( \kappa\,\alpha -1\right) \, {\cal{A}}_2=0  $}, which is solved by  ${\cal{A}}_2 = ({\alpha a+ \beta })^{\frac{1}{\kappa\alpha} -1}$.  Thus, we have two expressions for ${\cal{A}}_2$, both of which must be valid for all values of $a$. Their compatibility  implies $\frac{1}{\kappa\alpha} =2$, $\gamma=1$  and $\delta=0$.  Then ${\cal{A}}_2 = {\cal{A}}_1$. This then is an $N=1$ case.

\vspace*{0.25in}
Now let us look at the case of  { \{1,2,3\}}+\{4\}+\{5\}.

From, \{4\}, we get ${\cal{X}}_2 = \alpha_2 x + \beta_2. $
Since the  fifth term is constant, we have two choices: it is  either equal to zero or it is not. \begin{itemize}
\item
Let us first consider the case that  the fifth term is zero. As in the previous case, from  \{1,2,3\} we have
$    {\cal{A}}_1 \frac{ d \, {\cal{A}}_1}{da} \, \sim
{\cal{A}}_1 \sim \,  \, {\cal{A}}_2 \frac{ d \,
{\cal{A}}_2}{da}$; i.e.,   ${\cal{A}}_1= \alpha \,a+\beta$, and
that $\left(  {\cal{A}}_2 \right)^2 =
\gamma \left(  {\cal{A}}_1 \right)^2 +\delta$. But since the fifth term is zero, we also have  ${\cal{A}}_1\sim{\cal{A}}_2^{-1}$. These two conditions can only be met if both ${\cal{A}}_1$ and ${\cal{A}}_2$ are constants independent of $a$, and the problem reduces to the $N=1$ case.

\item
Now let us assume the fifth term is a non-zero $a$-dependent
constant.
 Then ${\cal{X}}_1\sim 1/{\cal{X}}_2.$ We have \{1,2,3\} and \{4\}
separately equal to $a$-dependent constants.  There are two possibilities for the fourth term:
it is either zero or a non-zero constant.  In the first instance, $\frac{ d \, {\cal{X}}_2}{dx}=0$, that makes  ${\cal{X}}_2$ a constant. This implies that  \{1,2,3\} is reducible, and hence will be considered later.  However, if the fourth term is not zero then
${\cal{X}}_2= \alpha x + \beta.$  This implies that
${\cal{X}}_1=\frac{\gamma}{ \alpha x + \beta}$.  Then the irreducible
three-term  \{1,2,3\}  implies:
\begin{eqnarray}
2{\cal{A}}_1 \,  \frac{ d \, {\cal{A}}_1}{da} \left(\frac{\gamma}{ \alpha x + \beta}\right)^2
- \,2 {\cal{A}}_1 \frac{ \gamma \alpha}{\left( \alpha x + \beta\right)^2}
+2{\cal{A}}_2 \,  \frac{ d \, {\cal{A}}_2}{da} (\alpha x + \beta) ^2
= M (a). \nonumber
\end{eqnarray}

This algebraic equation has a nontrivial solution for all values of $x$ only if $2{\cal{A}}_2 \,  \frac{ d \, {\cal{A}}_2}{da} =0$. However, that reduces  \{1,2,3\}  to  \{1,2\}+ \{3\} and hence will be considered later.
\end{itemize}

\item Terms \{1,2,4\} add up to an $a$-dependent constant.\\
I.e., $$    %
 2{\cal{A}}_1 \, \frac{ d \, {\cal{A}}_1}{da}\,   {\cal{X}}_1 ^{\,2}  \,
 -\,  2{\cal{A}}_1  \frac{ d \, {\cal{X}}_1}{dx} \,
 -\, 2 {\cal{A}}_2  \frac{ d \, {\cal{X}}_2}{dx} \,
= H(a).  %
$$
Then from the irreducibility of  { \{1,2,4\}}, ${\cal{A}}_1\sim{\cal{A}}_2$: an $N=1$ case.

\item Terms \{1,2,5\} add up to an $a$-dependent constant.
\begin{eqnarray}
 2{\cal{A}}_1 \, \frac{ d \, {\cal{A}}_1}{da}\,   {\cal{X}}_1 ^{\,2}  \,
 -\,  2{\cal{A}}_1  \frac{ d \, {\cal{X}}_1}{dx} \,
  + 2\, {\cal{X}}_1 {\cal{X}}_2 %
      \left( {\cal{A}}_1 \frac{ d \, {\cal{A}}_2}{da} +
             {\cal{A}}_2 \frac{ d \, {\cal{A}}_1}{da}
      \right)
= H(a).  %
\label{Term(12a)}
\end{eqnarray}
The remaining two terms could add up to a constant irreducibly or separately. Thus, there are two possibilities for this case: \{1,2,5\}+\{3,4\} or \{1,2,5\}+\{3\}+\{4\}. Before we analyze these cases, we note that from Eq. (\ref{Term(12a)}), we get
$ {\cal{A}}_1 \sim {\cal{A}}_1\frac{ d \, {\cal{A}}_1}{da} \sim{
     \left( {\cal{A}}_1 \frac{ d \, {\cal{A}}_2}{da} +
              {\cal{A}}_2 \frac{ d \, {\cal{A}}_1}{da}
      \right)}
$.
Thus, we have $ {\cal{A}}_1 = \alpha_1 a+\beta_1$.

Now, we first consider the case of \{1,2,5\}+\{3,4\}. From
 \{3,4\}, we have,
\begin{eqnarray}
{\cal{A}}_2  \frac{ d \, {\cal{A}}_2}{da} {\cal{X}}_2^{\,2} \,
 - {\cal{A}}_2  \frac{ d \, {\cal{X}}_2}{dx} = C(a)~.
\label{Term(12a1)}
\end{eqnarray}

From Eq. (\ref{Term(12a1)}), we get $ {\cal{A}}_2  \frac{ d \, {\cal{A}}_2}{da}  \sim  {\cal{A}}_2$, which gives ${\cal{A}}_2 = \alpha_2 a+\beta_2$.  Substituting expressions for $ {\cal{A}}_1$ and $ {\cal{A}}_2$ in $ {\cal{A}}_1 \sim{
     \left( {\cal{A}}_1 \frac{ d \, {\cal{A}}_2}{da} +
              {\cal{A}}_2 \frac{ d \, {\cal{A}}_1}{da}
      \right)} =  \alpha_2\, {\cal{A}}_1 + \alpha_1\, {\cal{A}}_2~
$ implies that  ${\cal{A}}_2\sim{\cal{A}}_1$, thus reducing this to an $N=1$ case.

Next we consider the case of \{1,2,5\}+\{3\}+\{4\}. From \{1,2,5\}, we still have  $ {\cal{A}}_1 = \alpha_1 a+\beta_1$ and ${\cal{A}}_1\frac{ d \, {\cal{A}}_1}{da} \sim{
     \left( {\cal{A}}_1 \frac{ d \, {\cal{A}}_2}{da} +
              {\cal{A}}_2 \frac{ d \, {\cal{A}}_1}{da}
      \right)} \sim  {\cal{A}}_1  \equiv \gamma_1 {\cal{A}}_1$. From \{3\} there are two possibilities. Either ${\cal{A}}_2$ is constant or ${\cal{X}}_2$ is constant. If ${\cal{A}}_2$ is constant, from \{1,2,5\} we find that ${\cal{A}}_1$ is also constant, and hence we have an $N=1$ case. Now we consider the case ${\cal{X}}_2=\gamma_2$ is constant. The irreducible  combination now leads to the differential equation
$$  \alpha_1 {\cal{X}}_1 ^{\,2}  \,
 - \frac{ d \, {\cal{X}}_1}{dx} \,
+\underbrace{\gamma_1 \gamma_2}_{\gamma} {\cal{X}}_1
= \frac{H(a)}{2{\cal{A}}_1}~,$$
whose solution is $\left(  \tanh x -\frac{\gamma}{2\alpha_1}\right)$.
To determine  ${\cal{A}}_2$, we start with the equation
${  \left( {\cal{A}}_1 \frac{ d \, {\cal{A}}_2}{da} + {\cal{A}}_2 \frac{ d \, {\cal{A}}_1}{da}
      \right)}  = \gamma_1 {\cal{A}}_1$; i.e.,
\begin{eqnarray}
(\alpha_1 a+\beta_1) \frac{ d \, {\cal{A}}_2}{da} + \alpha_1 {\cal{A}}_2
 = \gamma_1 (\alpha_1 a+\beta_1),
\end{eqnarray}
or
\begin{eqnarray}
 \frac{ d \, {\cal{A}}_2}{da} + \frac{\alpha_1 }{\alpha_1 a+\beta_1}~{\cal{A}}_2
- \gamma_1=0. \label{eq.26}
 \end{eqnarray}
Eq. (\ref{eq.26}) is solved by
\begin{eqnarray}
{\cal{A}}_2 =  \frac
{
\gamma_1 \left(\frac12  \alpha_1 a^2 +\beta_1 a     \right)+ \zeta_1
 }
 {\alpha_1 a+\beta_1}
 \end{eqnarray}
Solving for the superpotential $W(x,a) = {\cal{A}}_1 {\cal{X}}_1+{\cal{A}}_2 {\cal{X}}_2$, we get

\begin{eqnarray}
W(x,a) &=& ( {\alpha_1 a+\beta_1}) \tanh x + \frac{\zeta_1-  \frac{\gamma \, \beta_1^2}{2\alpha_1}} {\alpha_1 a+\beta_1} \nonumber \\
&=& A \tanh x + \frac{B} {A} ~,\nonumber
 \end{eqnarray}
where $A={\alpha_1 a+\beta_1}$ and $B={\zeta_1-  \frac{\gamma \, \beta_1^2}{2\alpha_1}} $.

\item
Terms \{1,3,4\} add up to an $a$-dependent constant.  This is the same as item (a).

\item Terms \{1,3,5\}  add up to an $a$-dependent constant.
\begin{eqnarray}    %
 2{\cal{A}}_1 \, \frac{ d \, {\cal{A}}_1}{da}\,   {\cal{X}}_1 ^{\,2}  \,
+    2 {\cal{A}}_2 \, \frac{ d \, {\cal{A}}_2}{da} \,   {\cal{X}}_2 ^{\,2} \,
  + 2\, {\cal{X}}_1 {\cal{X}}_2 %
      \left( {\cal{A}}_1 \frac{ d \, {\cal{A}}_2}{da} +
             {\cal{A}}_2 \frac{ d \, {\cal{A}}_1}{da}
      \right)
= H(a).  \label{Term135}%
\end{eqnarray}
Eq. (\ref{Term135}) implies that we have %
$ {\cal{A}}_1  \frac{ d \, {\cal{A}}_1}{da} \,
\sim \,  {\cal{A}}_2\frac{ d \, {\cal{A}}_2}{da} \,
\sim \,%
      \left( {\cal{A}}_1 \frac{ d \, {\cal{A}}_2}{da} +
             {\cal{A}}_2 \frac{ d \, {\cal{A}}_1}{da}
      \right)$.
${\cal{A}}_1  \frac{ d \, {\cal{A}}_1}{da} \, \sim \,  {\cal{A}}_2\frac{ d \, {\cal{A}}_2}{da}$ %
gives %
$\left( {\cal{A}}_1\right)^2 =\alpha_1 \left( {\cal{A}}_2\right)^2+\beta_1$. The proportionality condition $ {\cal{A}}_1  \frac{ d \, {\cal{A}}_1}{da} \,
\sim \,%
      \left( {\cal{A}}_1 \frac{ d \, {\cal{A}}_2}{da} +
             {\cal{A}}_2 \frac{ d \, {\cal{A}}_1}{da}
      \right)$ generates $\left( {\cal{A}}_1\right)^2 =\alpha_2 \left( {\cal{A}}_1 {\cal{A}}_2\right)+\beta_2$. If $\beta_1$ or   $\beta_2$ vanishes, we get  $ {\cal{A}}_1 \sim  {\cal{A}}_2$.  If $\beta_1$ and  $\beta_2$ are different from zero, these two algebraic equations for $ {\cal{A}}_1$ and $ {\cal{A}}_2$ determine them to be constants.  Thus, in both situations, we have an $N=1$ case.

\item Terms \{1,4,5\} add up to an $a$-dependent constant.\\
Let us first consider the case  \{1,4,5\} + \{2,3\} . Thus, we have
\begin{eqnarray}    %
 2{\cal{A}}_1 \, \frac{ d \, {\cal{A}}_1}{da}\,   {\cal{X}}_1 ^{\,2}  \,
 -\, 2 {\cal{A}}_2  \frac{ d \, {\cal{X}}_2}{dx} \,
  + 2\, {\cal{X}}_1 {\cal{X}}_2 %
      \left( {\cal{A}}_1 \frac{ d \, {\cal{A}}_2}{da} +
             {\cal{A}}_2 \frac{ d \, {\cal{A}}_1}{da}
      \right)
= H_1(a).  \label{Term145a}%
\end{eqnarray}
and
\begin{eqnarray}    %
 2{\cal{A}}_2 \, \frac{ d \, {\cal{A}}_2}{da}\,   {\cal{X}}_2 ^{\,2}  \,
 -\, 2 {\cal{A}}_1  \frac{ d \, {\cal{X}}_1}{dx} \,
= H_2(a).  \label{Term145b}%
\end{eqnarray}
Eqs. (\ref{Term145a}) and (\ref{Term145b}) imply
${\cal{A}}_1  \frac{ d \, {\cal{A}}_1}{da} \, \sim \,  {\cal{A}}_2 \sim \left( {\cal{A}}_1 \frac{ d \, {\cal{A}}_2}{da} +
{\cal{A}}_2 \frac{ d \, {\cal{A}}_1}{da}  \right)$, and
${\cal{A}}_2  \frac{ d \, {\cal{A}}_2}{da} \, \sim \,  {\cal{A}}_1$ respectively. Thus, we have
$\frac{ d \, {\cal{A}}_1}{da} \, \sim \, \frac{ {\cal{A}}_2}{{\cal{A}}_1} $ and
$\frac{ d \, {\cal{A}}_2}{da} \, \sim \, \frac{ {\cal{A}}_1}{{\cal{A}}_2} $.
From ${\cal{A}}_1  \frac{ d \, {\cal{A}}_1}{da} \sim \left( {\cal{A}}_1 \frac{ d \, {\cal{A}}_2}{da} + {\cal{A}}_2 \frac{ d \, {\cal{A}}_1}{da} \right)$, we get ${\cal{A}}_2 = \delta_1 {\cal{A}}_1 + \frac{\delta_2}{{\cal{A}}_1}$.
Substituting above expressions for $\frac{ d \, {\cal{A}}_1}{da}$, $\frac{ d \, {\cal{A}}_2}{da}$, and ${\cal{A}}_2 $ into ${\cal{A}}_2 \sim \left( {\cal{A}}_1 \frac{ d \, {\cal{A}}_2}{da} + {\cal{A}}_2 \frac{ d \, {\cal{A}}_1}{da}  \right)$, we get the following algebraic equation for ${\cal{A}}_1$:
$$\delta_1 {\cal{A}}_1 + \frac{\delta_2}{{\cal{A}}_1}  = \frac{\kappa_1 {\cal{A}}_1^2}{\delta_1 {\cal{A}}_1 + \frac{\delta_2}{{\cal{A}}_1} } + \frac{\left(  \delta_1 {\cal{A}}_1 + \frac{\delta_2}{{\cal{A}}_1}   \right)^2}{ {\cal{A}}_1}~~ ,$$
whose solution ${\cal{A}}_1$ equals a constant.  However, this then makes \{1,4,5\} irreducible, hence will be considered later.

Now consider the case  \{1,4,5\} + \{2\} + \{3\}. From \{3\}, we have ${\cal{A}}_2$ a constant because  ${\cal{X}}_2$ constant makes \{1,4,5\} reducible. Using this in \{1,4,5\}, we get ${\cal{A}}_1^2= \gamma_1\, a+\delta_1$ and ${\cal{A}}_1^2 = \gamma_2\, {\cal{A}}_1 {\cal{A}}_2+\delta_2$.  Since ${\cal{A}}_2$ is constant, from the second condition we get ${\cal{A}}_1$ constant. The problem thus reduces to the $N=1$ case.

\item Terms \{2,3,4\} add up to an $a$-dependent constant.
This is the same as (b).

\item Terms \{2,3,5\} add up to an $a$-dependent constant.
This is the same as (f).

\item Terms \{2,4,5\} add up to an $a$-dependent constant.
As stated earlier, the presence of  term \#2 and term \#4 in \{2,4,5\} implies ${\cal{A}}_1 \sim {\cal{A}}_2$ and the problem is immediately reduced to an $N=1$ case.
\item Terms \{3,4,5\} add up to an $a$-dependent constant.
This is the same as (c).

\end{enumerate}

\item As we set out to now consider all irreducible two term cases, let us recognize again that term \#1 in Eq. (\ref{coupled}) is equivalent to term \#3 and term \#2 is equivalent to term \#4.  There are several possibilities.
\begin{enumerate}
\item \begin{itemize}
\item  \{1,2\}+\{3,4\}+\{5\}\\
                  \{1,2\}+\{3,4\}+$\left( \{5\} =0\right)$

The first and second terms add to an $a$-dependent
constant:~
$2{\cal{A}}_1 \, \frac{ d \, {\cal{A}}_1}{da}\,   {\cal{X}}_1 ^{\,2}  \,
 -\,  2{\cal{A}}_1  \frac{ d \, {\cal{X}}_1}{dx} = H(a)$.
The proportionality of the coefficients on the left hand side
imply $\frac{ d \, {\cal{A}}_1}{da} = \alpha$.
Then %
${\cal{A}}_1 =\alpha a + \eta.$ Similarly, the constancy of the combination of the third and fourth terms gives
${\cal{A}}_2 =\alpha_2 a + \eta_2$ %

But from \{5\}, ${\cal{A}}_2\sim 1/{\cal{A}}_1$; i.e., $\alpha_2 a +
\eta_2\sim 1/ (\alpha_1 a + \eta_1)$.  For this to be true for all
values of $a$, $\alpha_1=\alpha_2=0$.  Then both ${\cal{A}}_1$ and
${\cal{A}}_2$ are $a$-independent constants: a special case of ${\cal{A}}_2\sim {\cal{A}}_1$, and thus an $N=1$ case.

 \vspace*{.25in}

                  \{1,2\}+\{3,4\}+$\left( \{5\} \neq 0\right)$
In this case, instead of ${\cal{A}}_2\sim 1/{\cal{A}}_1$,  from \{5\} we have ${\cal{X}}_2\sim 1/{\cal{X}}_1$.  The differential equations from \{1,2\} and \{3,4\} give respectively ${\cal{X}}_1 = \tanh x$ and ${\cal{X}}_2 = \coth x$.
As a result, the superpotential  for this case will be of the form:
\begin{eqnarray}
W(x,a) = (\alpha_1 a + \eta_1) \tanh \left( \gamma \,x \right) + (\alpha_2 a + \eta_2) \coth \left( \gamma \,x \right)~.
\end{eqnarray}
With the identification of $ (\alpha_1 a + \eta_1) = A$ and $(\alpha_2 a + \eta_2) =- B$, we get
$A \tanh r - B \coth r$.
This is the \emph{P\"oschl-Teller II} superpotential.
This also produces, with different constants, the \emph{P\"oschl-Teller I} superpotential \cite{Dutt}.

\item \{1,2\}+\{3,5\}+\{4\}\\
The constancy of the fourth term implies that $\frac{d{\cal{X}}_2}{dx} = \alpha$; i.e, ${\cal{X}}_2 = \alpha x +\beta$. We cannot consider the case of $\alpha=0$ here because that would make ${\cal{X}}_2$ constant, and thus \{3,5\}  would be reducible to \{3\} and \{5\}.  We will consider this later. From the irreducible \{3,5\}, we thus obtain
\begin{eqnarray}%
{\cal{A}}_2 \frac{ d \, {\cal{A}}_2}{da}\left(  {\cal{X}}_2 \right)^2+ {\cal{X}}_1{\cal{X}}_2\,
\frac{ d }{da}\left(  {\cal{A}}_1 {\cal{A}}_2\right) =D(a).\label{Eq.1}
 \end{eqnarray}
From this  we get
${\cal{A}}_2 \frac{ d \, {\cal{A}}_2}{da} \sim \frac{ d }{da}\left(  {\cal{A}}_1 {\cal{A}}_2\right)$, i.e., %
${\cal{A}}_1 = \theta_1 {\cal{A}}_2 + \frac{\theta_2}{ {\cal{A}}_2}. $
This leads to
\begin{eqnarray} \left(  {\cal{X}}_2 \right)^2 + \theta_1{\cal{X}}_1 {\cal{X}}_2  = \kappa.\label{Eq.2}\end{eqnarray}
For $ \kappa = 0$, we have an algebraic equation which reduces to ${\cal{X}}_1 \sim {\cal{X}}_2$; thus an $N=1$ case

For $ \kappa \neq  0$, from \{1,2\}, we have
${\cal{A}}_1 \frac{ d \, {\cal{A}}_1}{da}\left(  {\cal{X}}_1 \right)^2 - {\cal{A}}_1 \frac{d{\cal{X}}_1}{dx} = J(a),$ i.e., ${\cal{A}}_1 \frac{ d \, {\cal{A}}_1}{da} \sim {\cal{A}}_1 $. This gives
${\cal{A}}_1 =\gamma a + \nu$, which yields
$$\gamma \left( {\cal{X}}_1 \right)^2 -  \frac{d{\cal{X}}_1}{dx} = \lambda.$$
For $\lambda = 0$, we get ${\cal{X}}_1 = - \frac{\gamma }{x-x_0}$.  This is not compatible with the condition given in Eq. (\ref{Eq.2}) and ${\cal{X}}_2 = \alpha x + \beta$. Thus there is no solution.

For $\lambda \neq 0$, we get two cases again.  $\gamma$ could be equal to zero or a non-zero constant. For $\gamma = 0$,  \{1,2\} becomes reducible, and hence will be pursued later. For $\gamma \neq 0$, the solution for ${{\cal{X}}_1} $ will be a function of the type $\tan, \cot, \tanh, \, {\rm or}\, \coth$, none of which is compatible with ${{\cal{X}}_2} = \alpha x +\beta$ and Eq. (\ref{Eq.2}). Thus there is no solution.

\vspace*{0.15in}

\item  \{1,2\}+\{4,5\}+\{3\}\\
\hspace*{0.25in}
Constancy of the term \{3\} implies that either ${\cal{X}}_2=\xi$, a constant or the derivative of ${\cal{A}}_2$ zero, i.e., ${\cal{A}}_2 = \alpha_2$.
\begin{itemize}
\item If ${\cal{X}}_2$ is a constant, then \{4,5\} becomes reducible and will be discussed later.
\item
If ${\cal{A}}_2$ is constant, from \{4,5\} we have  ${\cal{A}}_1 = \alpha a +\beta$.   From \{1,2\}, we have
\begin{eqnarray}
    \alpha \, {\cal{X}}_1^2 -\frac{d{\cal{X}}_1}{dx} = \lambda. \label{eq_45a}
\end{eqnarray}
For $\lambda =0$, we get ${\cal{X}}_1 = -\frac{1}{\alpha\, x-\zeta_1}$
Substituting in the equation (from  \{4,5\})
\begin{eqnarray}\frac{ d \, {\cal{X}}_2}{dx}  - \alpha  {\cal{X}}_1 \, {\cal{X}}_2  =  \nu~,  \label{eq_45}\end{eqnarray}
and solving for $ {\cal{X}}_2$, we get
$$ {\cal{X}}_2 = \frac{ \nu \left( \frac\alpha2 x^2 - \zeta_1 x \right) +\zeta_2}{\alpha\, x-\zeta_1}.$$
The superpotential is then given by
$$\frac{ \alpha a +\beta +  \frac{\nu\xi \zeta_1}{2\alpha}  + \zeta_2}{\alpha\, x-\zeta_1}
-\frac{\xi\nu}{2\alpha} \left(  \alpha\, x-\zeta_1   \right)~,$$
which for $ \alpha a +\beta +  \frac{\nu\xi \zeta_1}{2\alpha}  + \zeta_2 = l+1$, $-\frac{\xi\nu}{2\alpha} = \frac\omega2$ and ${\alpha\, x-\zeta_1}=r$, becomes
\begin{eqnarray}
W(r,a) =\frac12\, \omega \,r - \frac{l+1}{r} ~,\nonumber
\end{eqnarray}
the \emph{three-dimensional harmonic oscillator}.

For $\lambda \neq 0$, we have ${\cal{X}}_1$ is of the form $\tan x$. Then solving for  ${\cal{X}}_2$ in Eq. (\ref{eq_45}), we get ${\cal{X}}_2$ of the form $\sec x$. Hence, the superpotential generated is of the form $$ W(x,A) = A \tan x + B\sec x   .$$
Depending on values of constants $\alpha, \nu$ and $\lambda$,  Eqs. (\ref{eq_45a}) and (\ref{eq_45}) also produce a shape invariant  superpotential: %
$$W(x,a) =A\coth x - B {\rm cosech}\,x .$$  This is the \emph{generalized P\"oschl-Teller} superpotential \cite{SUSYQM}. Similarly, Eqs. (\ref{eq_45a}) and (\ref{eq_45})  can also be used to generate  the \emph{Scarf I} $ (A\tan x - B {\rm sec}\,x$) and \emph{Scarf II} ($A\tanh x + B {\rm sech}\,x$)  superpotentials.
\end{itemize}

\item  \{1,2\}+\{3\}+\{4\}+\{5\}\\
From \{1,2\}, we have $ {\cal{A}}_1\frac{ d \, {\cal{A}}_1}{da} \,{\cal{X}}_1 ^{\,2}
 -\,{\cal{A}}_1  \frac{ d \, {\cal{X}}_1}{dx} \,
= M(a) $. This gives $\frac{ d \, {\cal{A}}_1}{da} =\alpha$,  constant; thus, ${\cal{A}}_1= \alpha a + \beta$.  For  $M(a) \neq 0$,  we then get $ {\cal{X}}_1 \sim \tanh$ or $ {\cal{X}}_1 \sim \coth$ or their trigonometric counterparts; i.e., $\tan$ and $\cot$. For  $M(a)  =  0$, the solution is $ {\cal{X}}_1  = - \frac{\alpha}{x}$.

For ${\cal{X}}_2$, let us turn to the remaining terms. The fifth term is either zero or a non-zero constant. We will consider them separately.

\{1,2\}+\{3\}+\{4\}+$\left( \{5\} =0\right)$\\  From \{3\}, either ${\cal{X}}_2$ or ${\cal{A}}_2$ is a constant.  If ${\cal{X}}_2$ is constant, with  ${\cal{A}}_1 \sim 1/{\cal{A}}_2$ from \{5\} =0 and $M(a) \neq 0$, we generate the following superpotentials: \emph{Rosen Morse I} $\left(- \frac B A - A\cot x \right)$, \emph{Rosen Morse II} $\left(\frac B A + A\tanh x\right)$, \emph{Eckart} $\left(\frac B A - A\coth x\right)$. For $M(a)  =  0$, we get instead $W(x,a) =  - \frac{\alpha (\alpha a + \beta)}{x}+\frac{\gamma}{\alpha a + \beta}$.  With the identifications $x\rightarrow r, ~\alpha (\alpha a + \beta)  \rightarrow (l+1)$ and $ \frac{\gamma}{(\alpha a + \beta)}= \frac{e^2}{2(l+1)}$, we obtain the \emph{Coulomb} superpotential.

 \{1,2\}+\{3\}+\{4\}+$\left( \{5\} \neq 0\right)$\\
From $\left( \{5\} \neq 0\right)$, we have ${\cal{X}}_1 \sim 1/{\cal{X}}_2$. Thus, ${\cal{X}}_2=$ constant is no longer a solution. From \{3\} and \{4\}, we find that ${\cal{X}}_2$ must be a linear function of $x$ and ${\cal{A}}_2$ must be a constant. Thus, only solution for  ${\cal{X}}_1$ that is compatible with ${\cal{X}}_1 \sim 1/{\cal{X}}_2$ and linearity of ${\cal{X}}_2$ is  $ {\cal{X}}_1 = - \frac{\alpha }{x}$.
Thus the superpotential generated is $W(x,a) = - \frac{\alpha (\alpha a + \beta)}{x}+\gamma x$. With the identification of \mbox{$x\rightarrow r, ~\alpha (\alpha a + \beta)  \rightarrow (l+1)$, and  $\gamma \rightarrow \frac12\, \omega$}, we get the superpotential of  the \emph{three-dimensional harmonic oscillator}:
\begin{eqnarray}
W(r,a) =\frac12\, \omega \,r - \frac{l+1}{r} ~,\nonumber
\end{eqnarray}
the same as \{1,2\}+\{4,5\}+\{3\}.

\end{itemize}
\item
\begin{itemize}
             \item  \{1,3\}+\{2,4\}+\{5\}

From \{2,4\},  this combination immediately reduces to an $N=1$ case.

\item
\{1,3\}+\{2\}+\{4\}+\{5\}.

Let us first consider \{1,3\}+\{2\}+\{4\}+$\left( \{5\} =0\right)$.

Since \{2\} and \{4\} are constants, we have  $ {\cal{X}}_1 = \alpha_1 \, x+\beta_1$ and ${\cal{X}}_2 = \alpha_2 \, x+\beta_2$. From \{1,3\}, we have
$$ (\alpha_1 \, x+\beta_1)^2+ \kappa (\alpha_2 \, x+\beta_2)^2 = \delta  .$$
The above equation is valid for all values of $x$, only if $\alpha_1 \, x+\beta_1 = \sqrt{-\kappa} \left( \alpha_2 \, x+\beta_2  \right)$; i.e., ${\cal{X}}_1 \sim {\cal{X}}_2$. Thus, we have a case of $N=1$.

This satisfies \{1,3\} only if $ {\cal{X}}_1$ and ${\cal{X}}_2$ are proportional. However, that again reduces the problem to the $N=1$ case.

\vspace*{.25in}
             \{1,3\}+\{2\}+\{4\}+$\left( \{5\} \neq 0\right)$,\\
   In this case since $\frac{d{\cal{X}}_1}{dx}$, $\frac{d{\cal{X}}_2}{dx}$ and ${\cal{X}}_1\,{\cal{X}}_2$
   are all constant, we get we get the trivial case of constant functions $ {\cal{X}}_1$ and ${\cal{X}}_2$.

\item  \{1,3\}+\{2,5\}+\{4\} \\
From \{4\},  we have ${\cal{X}}_2 =  \gamma x +\delta $.  Since terms 1 and 3 add irreducibly to a constant, we have $({\cal{X}}_1)^{\,2} +\kappa  ({\cal{X}}_2)^{\,2} =\nu$.  From this, we get ${\cal{X}}_1 =
\sqrt{ {\nu - \kappa\left(  \gamma x +\delta  \right)^2   }}$. Substituting these expressions of ${\cal{X}}_1$ and ${\cal{X}}_2$ in the combination \{2,5\}, we see that it is satisfied for a range of value of $x$ only if $\gamma=0.$ Thus, we get the trivial solution of ${\cal{X}}_2$ and ${\cal{X}}_2$ both constant.

\item  \{1,3\}+\{4,5\}+\{2\} is equivalent to the case \{1,3\}+\{2,5\}+\{4\} considered above; viz, the trivial solution.

\item  \{1,3\}+\{2\}+\{4\}+\{5\}

First consider \{1,3\}+\{2\}+\{4\}+(\{5\}=0). Since the sum of the first and third terms, $2{\cal{A}}_1 \, {\cal{X}}_1 ^{\,2} \, \frac{ d \, {\cal{A}}_1}{da} \, + 2{\cal{A}}_2 \,  {\cal{X}}_2 ^{\,2} \, \frac{ d \, {\cal{A}}_2}{da}=B(a)$, we have ${\cal{A}}_2^2 = \epsilon {\cal{A}}_1^2 + \gamma$. But from (\{5\}=0) we have ${\cal{A}}_2\sim 1/{\cal{A}}_1$.  Thus the only solution is that both ${\cal{A}}_1$ and ${\cal{A}}_2$ are $a$-independent constants, hence the special case of ${\cal{A}}_2\sim {\cal{A}}_1$; viz, the $N=1$ case.

Next we consider  \{1,3\}+\{2\}+\{4\}+(\{5\}$\neq$0).

From (\{5\}$\neq$0), we have ${\cal{X}}_2\sim 1/{\cal{X}}_1$. Terms \{2\} and \{4\} imply that ${\cal{X}}_1$ and ${\cal{X}}_2$ are linear functions of $x$. Compatibility among  \{2\}, \{4\}, and \{5\} thus requires that ${\cal{X}}_1$ and ${\cal{X}}_2$ both be constants, therefore we get the trivial solution again.
\end{itemize}%
\item %
\begin{itemize}
 \item  \{1,4\}+\{2,3\}+\{5\}\\
Let us first consider \{1,4\}+\{2,3\}+$\left( \{5\} =0\right)$. In this case, from \{1,4\} we have $\frac{ d \left( {\cal{A}}_1\right)^2}{da}  \sim {\cal{A}}_2$ and from $\left( \{5\} =0\right)$ we have ${\cal{A}}_1 \sim 1/{\cal{A}}_2$. The product of these two conditions give
${\cal{A}}_1 \frac{ d \left( {\cal{A}}_1\right)^2}{da}  \sim {\rm constant}$, i.e.,
$\frac{ d \left( {\cal{A}}_1\right)^3}{da} = \alpha, {\rm a~constant}.$ This gives ${\cal{A}}_1 = \left( \alpha a + \beta \right)^{1/3}$ and ${\cal{A}}_2 = \left( \alpha a + \beta \right)^{-1/3}$. However, from \{2,3\}, we also have $\frac{ d \left( {\cal{A}}_2\right)^2}{da}  \sim {\cal{A}}_1.$ This last constraint implies that we must have $\alpha=0$, i.e., ${\cal{A}}_1$ and ${\cal{A}}_2$ are constants; viz, the $N=1$ case.

Now let us consider \{1,4\}+\{2,3\}+$\left( \{5\} \neq 0\right)$.

From $\left( \{5\} \neq 0\right)$, we have  ${\cal{X}}_1 =\gamma /{\cal{X}}_2$. From \{2,3\} we get $\frac{d{\cal{X}}_1}{dx}+\alpha_1 {\cal{X}}_2^2 = \beta_1$, and from  \{1,4\} we have $\frac{d{\cal{X}}_2}{dx}+\alpha_2 {\cal{X}}_1^2 = \beta_2$. Differentiating ${\cal{X}}_1 =\gamma /{\cal{X}}_2$, we get $\frac{d{\cal{X}}_1}{dx} = - \gamma \frac{d{\cal{X}}_2}{dx}/\left({\cal{X}}_2\right)^2$. Eliminating ${\cal{X}}_1$ and its derivative from these two equations, we get an algebraic equation for ${\cal{X}}_2$, whose solution is ${\cal{X}}_2 = \delta$. This, along with ${\cal{X}}_1 =\gamma /{\cal{X}}_2$, reduces the problem to a trivial case of ${\cal{X}}_1$ and ${\cal{X}}_2$ equal to constants.

\item \{1,4\}+\{2,5\}+\{3\}\\
    Constancy of the term \{3\} implies that either ${\cal{X}}_2$ is constant or $\frac{ d \left( {\cal{A}}_2\right)^2}{da}=0$.
\begin{itemize}
\item
    First consider the case that ${\cal{X}}_2$ is constant. This gives $\{4\} =0$; i.e., \{1,4\} becomes a reducible case to be considered later.

\item
For the other possibility, $\frac{ d \left( {\cal{A}}_2\right)^2}{da}=0$, ${\cal{A}}_2$ is a constant. Then from \{1,4\}, we get $\frac{d}{da}\left( {\cal{A}}_1 \right)^2=  \theta_1$.
Which gives ${\cal{A}}_1 =  \sqrt{  \theta_1 a + \theta_2}.$  However, from \{2,5\} we get $ {\cal{A}}_1 \sim \frac{d}{da}{\cal{A}}_1$, i.e., $ {\cal{A}}_1 = \gamma_1 e^{\gamma_2\, a} + \gamma_3$. These two expressions are not compatible unless both $ {\cal{A}}_1$ and $ {\cal{A}}_2$ are constants, in which case $N=1$.
\end{itemize}

\item \{1,4\}+\{3,5\}+ \{2\}\\
\{2\} implies that $\frac{ d \,{\cal{X}}_1}{dx} = {\rm \,constant}$. So, ${\cal{X}}_1 = \alpha_1 x +\beta_1.$
\begin{itemize}
\item
$\alpha_1=0$.\\  Thus ${\cal{X}}_1 = \beta_1.$
This makes \{1,4\} reducible.

%

\item  $\alpha_1\neq0$.\\
From \{1,4\}, we have    ${\cal{X}}_2\sim (\alpha_1 x +\beta_1)^3 + \gamma$.

But \{3,5\} implies ${\cal{X}}_2^2\sim {\cal{X}}_1{\cal{X}}_2 +\delta.$
This expression is not compatible with the values of  ${\cal{X}}_1$ and ${\cal{X}}_2$ above, unless $ \alpha_1=0$; i.e.,  ${\cal{X}}_1$ and  ${\cal{X}}_2$ are constant, the trivial case once again.
\end{itemize}
\item \{1,4\}+\{2\}+\{3\}+\{5\}\\
As we saw above, constancy of the term \{3\} implies that either ${\cal{X}}_2$ is constant or ${\cal{A}}_2$ is constant, and ${\cal{X}}_2$  constant will not be considered, since it makes \{1,4\} reducible.
Thus ${\cal{A}}_2$ is constant.  For $\{5\}=0$, ${\cal{A}}_1\sim 1/{\cal{A}}_2=$ a constant, thus reducing to an $N=1$ case.
For $\{5\}\neq 0$, we have ${\cal{X}}_1\sim 1/{\cal{X}}_2$. Term \{2\} constant means that ${\cal{X}}_1=\alpha_1 x + \beta_1$, hence  ${\cal{X}}_2 \sim 1/(\alpha_1 x + \beta_1)$.
But from \{1,4\}, $\frac{ d \, {\cal{X}}_2}{dx}=\epsilon {\cal{X}}_1^2+ \delta$.  So ${\cal{X}}_2=\epsilon (\alpha_1 x + \beta_1)^3/(3\alpha_1)+ \delta x$.  These are incompatible expressions; therefore, there is no solution.
           \end{itemize}
\item  \begin{itemize}
 \item  \{1,5\}+\{2,3\}+\{4\}.  This is equivalent to \{1,4\} + \{3,5\} + \{2\} analyzed earlier.
 \item  \{1,5\}+\{2,4\}+\{3\}.
The presence of  \{2,4\} immediately implies an $N=1$ case.

 \item  \{1,5\}+\{3,4\}+\{2\} is equivalent to \{1,2\} + \{3,5\} + \{4\} analyzed earlier.

 \item  \{1,5\}+\{2\}+\{3\}+\{4\}
 From \{2\} and \{4\}, we have ${\cal{X}}_1 = \alpha_1 x +\beta_1$ and ${\cal{X}}_2 = \alpha_2 x +\beta_2$. Then from \{3\}, we have two possibilities: either ${\cal{X}}_2$ is a constant or ${\cal{A}}_2$ is constant.
 \begin{itemize}
 \item
 We first consider the case ${\cal{A}}_2$ constant. This implies, from \{1,5\}, ${\cal{A}}_1$ is  constant as well. This is an $N=1$ case.
  \item
 Now we consider the case ${\cal{X}}_2=\beta_2$, a constant. Since we have ${\cal{X}}_1 = \alpha_1 x +\beta_1$, from \{1,5\} we get
 $${\cal{A}}_1\,
 \frac{d{\cal{A}}_1}{da}
 \left( \alpha_1 x +\beta_1 \right)^2+\beta_2 \left( \alpha_1 x +\beta_1 \right)
 \frac{d{\cal{A}}_1 {\cal{A}}_2}{da}
 = J(a).
 $$
This equation can be satisfied for all values of $x$, only if $\alpha_1 =0$ , in which case ${\cal{X}}_1$ is also constant; viz., a trivial solution, or if  ${\cal{A}}_1$ and ${\cal{A}}_2$ are independent of $a$, reducing this to an $N=1$ case.
 \end{itemize}
\end{itemize}
\item
\begin{itemize}
 \item  \{2,4\}+\{1\}+\{3\}+\{5\}.~~   \{2,4\} gives ${\cal{A}}_1 \sim{\cal{A}}_2$, an $N=1$ case.
           \end{itemize}
\item  \begin{itemize}
 \item  \{2,5\}+\{1\}+\{3\}+\{4\}
\end{itemize}
Since term \{1\} is constant, we have two choices. It is either equal to zero or a non-zero constant.

\begin{itemize}
\item
If it is non-zero, that would imply that ${\cal{X}}_1$ is a constant. This reduces  \{2,5\} to  \{2\} +\{5\} and will be considered later.

\item
On the other hand, if term \{1\} is zero, we must have ${\cal{A}}_1$ a constant (since ${\cal{X}}_1=0$ is not acceptable.)  So we choose ${\cal{A}}_1= \eta$. From \{2,5\} we then get $\frac{d{\cal{A}}_2}{da} =\alpha$, so ${\cal{A}}_2=\alpha a + \beta$. Further, from  \{3\} we have $ {\cal{X}}_2=\delta$, a constant , since $ {\cal{A}}_2$ is not. This then gives
$
\left(
           \frac{ d \, {\cal{X}}_1}{dx}  - \alpha \delta {\cal{X}}_1 %
\right)
= \mu$.  %
The solution of this equation is given by ${\cal{X}}_1 = \Lambda e^{(\alpha \delta \,x)} - \frac{\mu}{\alpha \delta }$.
Thus the resulting superpotential
$W(x,a) = {\cal{A}}_1{\cal{X}}_1+ {\cal{A}}_2{\cal{X}}_2 $   is given by
\begin{eqnarray}
W(x,A) = A- B e^{-\nu\,x} , \label{Morse}
\end{eqnarray}
where $A=(\alpha a + \beta) -\frac{\eta\mu}{\alpha \delta } , ~B=\eta, ~\nu = \alpha \delta  .$   For $\alpha<0$,  this gives the \emph{Morse} superpotential.
\end{itemize}
 \item \{1\}+\{2\}+\{3\}+\{4\}+\{5\}.  \\
 Let us consider various possibilities. \\
  (\{1\} = 0) implies ${\cal{A}}_1$ is a constant.
  (\{1\}$\neq$0) implies ${\cal{X}}_1$ is a constant.
  (\{3\} = 0) implies ${\cal{A}}_2$ is a constant.
  (\{3\}$\neq$0) implies ${\cal{X}}_2$ is a constant.
So, if both   (\{1\} = 0) and   (\{3\} = 0), we have ${\cal{A}}_1 \sim {\cal{A}}_2$, hence $N=1.$
 Similarly, if both (\{1\}$\neq$0) and (\{3\}$\neq$0), we have ${\cal{X}}_1$ and ${\cal{X}}_2$ equal constants, a trivial case. Thus, the only case that we need to consider is   (\{1\}$\neq$0) and   (\{3\} = 0). This equivalent to (\{1\}=0) and   (\{3\} $\neq$ 0).  In this case, we have ${\cal{A}}_1=\nu_1$ and ${\cal{X}}_2=\beta_2$. Then  \{5\} yields $\nu_1 \beta_2 {\cal{X}}_1 \frac{{\cal{A}}_2}{da}$. If (\{5\} = 0), we have ${\cal{A}}_2=\nu_2$; i.e., ${\cal{A}}_1 \sim {\cal{A}}_2$, an $N=1$ case.  If (\{5\} $\neq$ 0), ${\cal{X}}_1=\beta_1$ and hence a trivial case where both ${\cal{X}}_1$  and ${\cal{X}}_2$ are constants.

\vspace*{.25in}

This exhausts all the possibilities for the $N=2$ ansatz.

In particular, we have now obtained all known translational shape invariant superpotentials.  We have found no new ones, using the $N=1$ and $N=2$ ansatz. Table I lists these superpotentials, as well as one of the combinations of irreducible terms from which they were obtained~\footnote{As we have seen, there are multiple combinations of irreducible terms which obtain the same superpotential.}.

\end{enumerate}

\end{enumerate}
\vspace*{.75in}

\begin{center}
\begin{tabular}{||l|l|l|l||}  \hline
Name  &  superpotential   & N & Combination \\ \hline
Harmonic Oscillator &  $\frac12 \omega x$ &  1 & \\
  \hline
Coulomb   &  $\frac{e^2}{2(l+1)} - \frac{l+1}{r}$& 2&\{1,2\}+\{3\}+\{4\}+$\left( \{5\} =0\right)$ \\
    \hline
Three dim. harmonic oscillator  & $\frac12 \omega r - \frac{l+1}{r}$ & 2&\{1,2\}+\{3\}+\{4\}+$\left( \{5\} \neq 0\right)$  \\
 \hline
Morse &$A-Be^{-x}$ & 2&\{2,5\}+\{1\}+\{3\}+\{4\}  \\
 \hline
P\"oschl-Teller I &$A \tan r - B \cot r$ & 2& \{1,2\}+\{3,4\}+$\left( \{5\} \neq 0\right)$  \\
 \hline
P\"oschl-Teller II &$A \tanh r - B \coth r$& 2& \{1,2\}+\{3,4\}+$\left( \{5\} \neq 0\right)$  \\
 \hline
Rosen-Morse I &$-A\cot x - \frac{B}{A}$ & 2& \{1,2\}+\{3\}+\{4\}+$\left( \{5\} =0\right)$ \\
 \hline
Rosen-Morse II &$A\tanh x + \frac{B}{A}$ & 2& \{1,2\}+\{3\}+\{4\}+$\left( \{5\} =0\right)$\\
 \hline
Eckart &$-A\coth x + \frac{B}{A}$ & 2& \{1,2\}+\{3\}+\{4\}+$\left( \{5\} =0\right)$\\
 \hline
{Scarf I} &$ A\tan x + B {\rm sec}\,x$ & 2&\{1,2\}+\{4,5\}+\{3\}  \\
 \hline
{Scarf II}  &$A\tanh x + B {\rm sech}\,x$& 2&\{1,2\}+\{4,5\}+\{3\}  \\
 \hline
Generalized P\"oschl-Teller &$A\coth x- B {\rm cosech}\,x $ & 2&\{1,2\}+\{4,5\}+\{3\}  \\
 \hline
\end{tabular}

\vspace*{.25in}
\begin{center}
{Table 1}
\end{center}

\end{center}
\newpage
\vspace*{.25in}\noindent
{\bf Conclusions}
\vspace*{.2in}

Until now, finding shape invariant superpotentials was left to excellent intuitions or trial and error.  The procedure described here is an {\it ab initio} method for generating superpotentials that are shape invariant, and hence exactly solvable.

We have transformed the translational shape invariance condition into a nonlinear partial differential equation.  With this step, we have transformed a non-local  difference-differential equation into a differential equation which must be satisfied by all translational shape invariant superpotentials. We have also constructed a variant of the ``separation of variables" method to find solutions of this nonlinear partial differential equation, and have found that the list of solutions generated includes all shape invariant superpotentials, as given in Ref.\cite{Dutt}.

\vspace*{.15in}\noindent

This work can be extended in several directions. On the one hand, other forms of shape invariance can
be investigated, to ascertain whether a similar transformation of difference equation into differential equation is possible. In particular, while we have not considered the case of multiplicative shape invariance, we believe that this work can be extended in that direction.

\vspace*{.15in}
On the other hand, in an earlier work \cite{AGJVMUS} we argued, based on group theoretical considerations,  that all of the known cases of translational shape invariance are the only ones which can exist. We have verified this claim for $N = 1,2$.  However, it remains to be checked whether our claim holds for larger values of $N$.

\newpage
{\Large
\begin{center}
{\bf Appendix}\\
Linear dependence of coefficients $F_i$ for an irreducible set
\end{center}
}
The nonlinear partial differential equation that we want to solve is given by
\begin{eqnarray}
   \underbrace{2{\cal{A}}_1 \, \frac{ d \, {\cal{A}}_1}{da}\,   {\cal{X}}_1 ^{\,2} }_{\rm Term \# 1}
 - \underbrace{2{\cal{A}}_1  \frac{ d \, {\cal{X}}_1}{dx}}_{\rm Term \# 2}
+  \underbrace{2 {\cal{A}}_2 \, \frac{ d \, {\cal{A}}_2}{da} \,   {\cal{X}}_2 ^{\,2}}_{\rm Term \# 3}
-   \underbrace{2 {\cal{A}}_2  \frac{ d \, {\cal{X}}_2}{dx}}_{\rm Term \# 4}
+   \underbrace{2\, {\cal{X}}_1 {\cal{X}}_2 %
      \left( {\cal{A}}_1 \frac{ d \, {\cal{A}}_2}{da} +
{\cal{A}}_2 \frac{ d \, {\cal{A}}_1}{da}
      \right)}_{\rm Term \# 5}
  = -  \underbrace{ \, \frac{d g(a)}{d a}}_{\rm Term \# 6}
\label{Linear1} \end{eqnarray}
This is of the form
\begin{eqnarray}
\sum_i^6 F_i(a) G_i(x)=0
\label{Linear2} \end{eqnarray}
where
\begin{equation}
\begin{array}{ll}
F_1 \equiv 2\,   {\cal{A}}_1 \, \frac{ d \, {\cal{A}}_1}{da};~~~~~~~~~~~~~~~~~~~~~~~~~~~~
&G_1 \equiv \left( {\cal{X}}_1 \right)^{2}\\%
F_2 \equiv -\,  2 {\cal{A}}_1   ;
&G_2 \equiv \frac{ d \, {\cal{X}}_1}{dx} \\%
F_3 \equiv 2\,   {\cal{A}}_2 \, \frac{ d \, {\cal{A}}_2}{da} ;
&G_3 \equiv \left( {\cal{X}}_2 \right)^{2}\\%
F_4 \equiv -\,  2 {\cal{A}}_2   ;
&G_4 \equiv \frac{ d \, {\cal{X}}_2}{dx} \\%
F_5 \equiv 2\,    \left( {\cal{A}}_1 \frac{ d \, {\cal{A}}_2}{da} +
{\cal{A}}_2 \frac{ d \, {\cal{A}}_1}{da}
      \right) ;
&G_5 \equiv {\cal{X}}_1 {\cal{X}}_2\\%
F_6 \equiv  dg/da   ;
&G_6 \equiv 1
\end{array}
\end{equation}

\vspace*{.2in}
Eq. (\ref{Linear2}) shows that not all $F_i$ are linearly independent. In particular, it states that at most five $F_i$ can be linearly independent and the dimensionality of the space they would span could at most be five. We will show that if the expression irreducibly adds up to a constant, the space spanned by $F_i$ would have the dimensionality of one; i.e., all  $F_i$ must be proportional to each other.

Let us first consider an expression consisting of just two elements: $\left\{ F_1G_1,  F_2G_2 \right\}$, that add up to a constant irreducibly; i.e., $F_1(a)G_1(x)+ F_2(a)G_2(x)= H(a)$. The irreducibility for this case implies that $F_1(a)G_1(x)$ and $ F_2(a)G_2 (x)$ cannot be separately constant.  This implies that neither $G_1(x)$ nor $G_2(x)$ can be $x$-independent constants, and $F_1(a)$ and $F_2(a)$ cannot be equal to zero, otherwise the two terms would be reducible to one.

Dividing the above equality by $H(a)$ and defining ${\cal F}_1(a) = \frac{F_1(a)}{H(a)},~{\cal F}_2(a) = \frac{F_2(a)}{H(a)}$,
\begin{eqnarray}
{\cal F}_1(a)G_1(x)+ {\cal F}_2(a)G_2 (x)= 1 .\label{Two-term}
\end{eqnarray}
Since $x$ and $a$ are real variables, the above expression must be valid for infinitely many values of both $x$ and $a$. Let us consider $a_1$ and $a_2$, two arbitrarily chosen values of $a$. For them, we have
\begin{eqnarray}
{\cal F}_1(a_1)G_1(x)+ {\cal F}_2(a_1)G_2 (x)&=& 1 \nonumber\\
{\cal F}_1(a_2)G_1(x)+ {\cal F}_2(a_2)G_2 (x)&=& 1 .\label{Two-term2}
\end{eqnarray}
Imagine a plane where $G_1(x)$ and $G_2(x)$ are $x$ and $y$ coordinates respectively. Each of the equalities expressed in Eq. (\ref{Two-term2}) are represented by a line on this $G_1$-$G_2$ plane. If these two lines intersect, we will have a solution for both $G_1(x)$ and $G_2(x)$; in other words, each would be a determined constant. However, our hypothesis of irreducibility states that we should not be able to determine such solution; i.e. Eq. (\ref{Two-term2}) should be invertible. This implies that we must have the lines parallel; i.e., both must have the same slope:
\begin{eqnarray}
\frac{ {\cal F}_2(a_1)}{ {\cal F}_1(a_1)} = \frac{ {\cal F}_2(a_2)}{ {\cal F}_1(a_2)}
\end{eqnarray}
In other words, the ratio $\frac{ {\cal F}_2(a)}{ {\cal F}_1(a)}$ is independent of the argument $a$, and hence must be equal to an $a$-independent constant.  This implies $F_1$ and $F_2$ are proportional to $H.$

Let us now consider an example that consists of three terms that irreducibly add up to a constant. So, we have
\begin{eqnarray}
{\cal F}_1(a)G_1(x)+ {\cal F}_2(a)G_2 (x) + {\cal F}_3(a)G_3 (x)= 1 .\label{Three-term}
\end{eqnarray}
Again, we can define ${\cal F}_i(a) = \frac{F_i(a)}{H(a)}$ for $i=1,2,3$, and obtain the following set of three equations for three arbitrary values of $a$:
\begin{eqnarray}
{\cal F}_1(a_1)G_1(x)+ {\cal F}_2(a_1)G_2 (x)+ {\cal F}_3(a_1)G_3 (x)&=& 1 \nonumber\\
{\cal F}_1(a_2)G_1(x)+ {\cal F}_2(a_2)G_2 (x)+ {\cal F}_3(a_2)G_3 (x)&=& 1 \nonumber\\
{\cal F}_1(a_3)G_1(x)+ {\cal F}_2(a_3)G_2 (x)+ {\cal F}_3(a_3)G_3 (x)&=& 1.\label{Three-term2}
\end{eqnarray}
In this case, consider a three dimensional space where $G_i(x)$ ($i=1,2,3$) are along coordinate axes. Eqs. (\ref{Three-term2}) represent three planes. If these planes intersected, the functions $G_i(x)$ will get determined and our expression would reduce to two-term or single-term expression. Since we stipulated that our three-term expression was irreducible, these planes must all be parallel. This implies that the direction ratios of their normals; i.e.,
$
    \left\{
        { {\cal F}_1(a_i) } ,
        { {\cal F}_2(a_i) } ,
        { {\cal F}_3(a_i) }
    \right\}
$ must be proportional:
$$
        \frac{ {\cal F}_1(a_1) }{ {\cal F}_1(a_2) } =
        \frac{ {\cal F}_2(a_1) }{ {\cal F}_2(a_2) } =
        \frac{ {\cal F}_3(a_1) }{ {\cal F}_3(a_2) }
    ~~~~~~~~{\rm etc.}
$$
must all be equal for different values of $a_i$.
This equation can be rearranged to yield
\begin{eqnarray}
\frac{ {\cal F}_1(a_i) }{ {\cal F}_2(a_i) }=\frac{ {\cal F}_1(a_j) }{ {\cal F}_2(a_j)} ~;~~~~~~~~
\frac{ {\cal F}_1(a_i) }{ {\cal F}_3(a_i) }=\frac{ {\cal F}_1(a_j) }{ {\cal F}_3(a_j)}~~.
\end{eqnarray}
This means that ratios such as $\frac{ {\cal F}_1(a_i) }{ {\cal F}_2(a_i) }$ do not depend on $a$. Thus ${\cal F}_1(a)  \sim {\cal F}_2(a)  \sim {\cal F}_3(a) .$  This implies $F_1$, $F_2$, and $F_3$ are proportional to $H.$

The method we used for two terms is identical to the one we used for the three-term case and can therefore be similarly extended to higher dimensional spaces for any of the four- or five-term irreducible expressions we have used in this paper.
\end{document}